\documentclass[10pt]{article}
\usepackage{latexsym,graphicx}
\usepackage{cite}
\usepackage{amsmath}
\usepackage{amsfonts}
\usepackage{amssymb}
\newcommand{\be}{\begin{equation}}
\newcommand{\ee}{\end{equation}}
\usepackage[left=2.5cm,top=2.5cm,right=2.5cm,bottom=1.5cm]{geometry}

\catcode `\@=11
\catcode `\@=12
\usepackage{graphicx}
\usepackage{epstopdf}
\usepackage{float}
\usepackage{cite}
\DeclareGraphicsExtensions{.eps}
\begin{document}
\begin{center}
\large{\bf{ Cosmological models of generalized ghost pilgrim dark energy (GGPDE) in the gravitation theory of Saez-Ballester }} \\
\vspace{10mm}
\normalsize{PRIYANKA GARG$^1$, ARCHANA DIXIT$^2$, ANIRUDH PRADHAN$^3$} \\
\vspace{5mm}
\normalsize{$^{1,2,3}$Department of Mathematics, Institute of Applied Sciences \& Humanities, GLA University,\\ 
Mathura -281 406, Uttar Pradesh, India \\
\vspace{2mm}
$^1$E-mail:pri.19aug@gmail.com \\
\vspace{2mm}
$^2$E-mail: archana.dixit@gla.ac.in  \\
\vspace{2mm}
$^3$E-mail:pradhan.anirudh@gmail.com }\\
\end{center}
\vspace{10mm}
\begin{abstract}
We are studying the mechanism of the cosmic model in the presence of GGPDE 
and matter in LRS Bianchi type-I space-time by the utilization of new holographic DE in Saez-Ballester theory. Here we 
discuss all the data for three scenarios, first is supernovae type Ia union data, second is SN Ia data in combination 
with BAO and CMB observations and third is combination with OHD and JLA observations. From this, we get a model of our 
universe, where its transit state from deceleration to acceleration phase. Here we have observed that the results yielded by 
cosmological parameters like $\rho$ (energy density), EoS (equation of state), squared speed of sound $(v_s^2)$, $(\omega_D-\omega_D^{'})$ 
and $(r-s)$ plane is compatible with the recent observational data. The $(\omega_D-\omega_D^{'})$ trajectories in both thawing 
and freezing regions and the correspondence of the quintessence field with GGPD dark energy are discussed. Some physical 
aspects of the GGPDE models are also highlighted.
\end{abstract}
 \smallskip
 Keywords. ~~ LRS Bianchi type-I metric; Ghost Pilgrim dark energy;
 time-dependent deceleration parameter; Saez-Ballester theory. \\
 \smallskip
 PACS Nos ~~ 98.80.Jk; 95.36.+x; 98.80.-k \\

\section{Introduction}
Nowadays it is accepted by almost every scientist in cosmology and relativity field that the cosmos is expanding till 1998, and 
the expansion rate is decreasing, because of gravity following up on the matter. In 1998 and following years, many groups of 
astronomers \cite{ref1,ref2,ref3,ref4,ref5,ref6}, in their search for the estimation of the universe expansion rate have 
observed some surprising results. These groups, in view of  SN Ia observations, have estimated the separations and predicted 
the accelerated expansion of the cosmos, which is probably going to continue forever. It is anticipated by these observations 
that something is responsible for this acceleration. To answer this, the cosmological constant $\Lambda$, which can be the 
best candidate for dark energy (DE). Researchers \cite{ref7,ref8} have discussed locally rotationally symmetric ghost pilgrim 
dark energy cosmological models. In the Saez-Ballester theory of gravitation Bianchi type-II, VIII, and IX dark energy models have 
been discussed by \cite{ref9}. Many authors  \cite{ref3,ref10}, have observed the cosmological observation that the expansion 
of the universe is accelerating.\\  

To rectify the problem of DE, various DE models such as phantom, tachyon, quintessence, and Chaplying gas have been proposed by
cosmologists from time to time. The cosmology scientist usually accepts that a type of repulsive force that
works as anti-gravity generated somewhere in the range of $7$ billion years back is the reason for speeding up the expansion of
the universe. This obscure physical entity is named as `dark energy'. Along with this, various researchers considered many alternative
candidates as a possible source of dark energy in different scenario \cite{ref11,ref12,ref13,ref14} in the Saez-Ballester theory.
Higher-dimensional cosmic models have been discussed by \cite{ref15,ref16,ref17,ref18}.\\

 Other than these research, a few cosmologists have studied holographic DE models as other options to DE causing late-time speeding 
 expansion of the universe \cite{ref19,ref20,ref21,ref22,ref23,ref24,ref25,ref26,ref27,ref28,ref29,ref30}. The hypothesis, that 
 phantom-like DE has an adequate resistant force to cause the formation of black hole in our universe later, is the base of the 
 proposal of pilgrim dark energy. Depending on the assumption of the generalized ghost version of the pilgrim dark energy this 
 phenomenon has been explored. We found that the parameter (EoS) varies in both quintessence and phantom regions. Here we use 
 linearly varying deceleration parameter (LVDP) to test the GGPD dark energy model in the SB scalar-tensor theory of gravity.
 We discuss the EoS parameter, squared speed of sound $(v_s^2)$,  $(\omega_D-\omega_D^{'})$ trajectories and $(r-s)$ planes to analyze the 
 results \cite{ref31}. Nowadays, due to the global property of the Universe the holographic dark energy has attracted much attention. 
 This model depends on the holographic principle \cite{ref32,ref33,ref34,ref35,ref36,ref37,ref38,ref39}.  \\
 
 Furthermore, a model of dark energy named  PDE (pilgrim dark energy) has been introduced by \cite{ref40}, and it is based on the fact 
that the strong repulsive force of dark energy can avoid the formation of the black hole. Instead of other types of DE, the 
phantom DE can play an important role. This argument coincides with the concept given by \cite{ref41}  which says that the mass 
of black holes tends to zero in the phantom energy universe approaching the Big Rip. $\rho_{\Lambda}$ (energy density) of GGPDE is 
discussed and given in \cite{ref42,ref43}. Cosmological model in $f(R,T)$ theory of gravity are analyzed in \cite{ref44, ref45}. 
The detail expression for it is given by Eq. (\ref{13}) in Sect. $3$. \\

In the non-flat FRW universe has studied the analysis of GGPDE in \cite{ref46}. The GGPDE of dark energy is also investigated in \cite{ref47}.
In general relativity, some Bianchi type GGPDE models have been studied by  \cite{ref48, ref49}. Currently \cite{ref50} have studied 
Bianchi-type-$VI_0$ GGPDE model. Researcher \cite{ref51} have discussed the cosmological implication 
of interacting PDE models (with Hubble, Granda-Oliveros and generalized ghost cutoffs) $\Lambda$CDM in fractal cosmology by taking the 
flat universe. The GPDE model in $F(T,T_{G})$ gravity with flat FRW universe have studied in \cite{ref52}. Interacting GGDE in a non-flat 
universe has been discussed by \cite{ref53} \\ 

In this paper, our aim is to investigate the LRS Bianchi type-I GGPDE model in the Saez-Ballester theory of gravitation. 
The paper has the following structure. Sect. $2$, presents deviation of field equations in Saez-Ballester's theory. In Sect. $3$, 
we obtained the metric and field equations for Bianchi type-I space times. In Sect. $2$, we describe calculations of some other 
physical and kinematic parameters. Interpretation of the results obtained is written in Sect. $3$. In Sect. $4$, we deal the the 
stability analysis of the solution. State-finder p ($r,s$) have been obtained in Sect. $5$. Sound speed 
is discussed in Sect. $6$. The analysis of $\omega_D -\omega^{'}_D$ in Sect. $7$. Correspondence of quintessence
field in Sect. $8$. In the last part Sect. $9$, conclusion and the description are given.

\section {Explicit field equations in Saez-Ballester theory}
The field equations founded by Saez-Ballester \cite{ref1} for the combined scalar
 and tensor fields are:

\begin{equation}
\label{1}
R_{ij} -\frac{1}{2} R g_{ij} - w \phi^n(\phi_{,i} \phi_{,j}-\frac{1}{2} g_{ij} \phi_{,k}\phi^{,k} )= -(T_{ij}+\bar{T}_{ij}),
\end{equation}

and the equations which are satisfied by $\phi$ (scalar field) :
\begin{equation}
\label{2}
2 \phi^n {\phi_{;i}^{,i}} +n\phi^{n-1}\phi_{,k}\phi^{,k} =0.
\end{equation}
In addition the conservation equation is
\begin{equation}
\label{3}
(T_{ij}+\overline{T}_{ij})_{;j} =0.
\end{equation}
Here, $R$ is the Ricci scalar, $R_{ij}$ is the Ricci tensor, $n$ and $w$ are arbitrary dimensionless constants and $8\pi G$ = C = $1$ in 
the relativistic units. \\

 Now, we define $T_{ij}$ and $\overline{T}_{ij}$  which are energy-momentum tensors for matter and holographic dark energy such as 
\begin{equation}
\label{4}
T_{ij} = \rho_m u_i u_j.
\end{equation}
\begin{equation}
\label{5}
\overline{T}_{ij} = (\rho_\Lambda + p_\Lambda)u_i u_j  - p_\Lambda g_{ij}.
\end{equation}
Here $\rho_\Lambda$ is the holographic dark energy, $\rho_m$ is the energy densities of matter and $p_\Lambda$ is the pressure of holographic 
dark energy.\\

Now LRS Bianchi type-I metric takes the form 
 \begin{equation}
 \label{6}
 ds^{2}=dt^{2}-a_{1}^{2}dx^{2} - a_{2}^{2} (dy^{2}+ dz^{2}),
 \end{equation}
 where $a_{1}(t)$ and $a_{2}(t)$ are time dependent functions. \\
 
 Now using Eqs. (\ref{1}), (\ref{2}), (\ref{4}), (\ref{5}) and co moving coordinates for the metric (\ref{6}) form the following equations 
\begin{equation}
\label{7}
\frac{2\ddot{a_{2}}}{a_{2}}+\frac{\dot{a_{2}}^2}{a_{2}^2}-
\frac{w}{2}{\phi^n} {\dot{\phi^2}}= -p_{\Lambda},
 \end{equation}
\begin{equation}
\label{8}
\frac{\ddot{a_{1}}}{a_{1}}+ \frac{\ddot{a_{2}}}{a_{2}}+
\frac{\dot{a_{1}}\dot{a_{2}}}{a_{1}a_{2}} - \frac{w}{2}{\phi^n} {\dot{\phi^2}}= -p_{\Lambda},
\end{equation}

\begin{equation}
\label{9}
\frac{2\dot{a_{1}}\dot{a_{2}}}{a_{1}a_{2}}+\frac{\dot{a_{2}}^2}{a_{2}^2}+ \frac{w}{2}{\phi^n} {\dot{\phi^2}} = \rho_m + \rho_\Lambda,
\end{equation}

\begin{equation}
\label{10}
\ddot{\phi} + \dot{\phi}\left(\frac{\dot a_{1}}{a_{1}} + 2\frac{\dot a_{2}}{a_{2}}\right) + \frac{n}{2}~\frac{\dot{\phi^2}}{\phi} = 0.
\end{equation}
Here  overhead dot indicates differentiation with respect to $t$.
Here the EoS parameter of dark energy is ${p_\Lambda }= {\omega_{\Lambda} }{\rho_\Lambda}$. \\

The conservation equation of the matter and dark energy is
\begin{equation}
\label{11}
\dot{\rho_{m}} + \dot{\rho_{\Lambda}} + \left(\frac{\dot a_{1}}{a_{1}} + 2\frac{\dot a_{2}}{a_{2}}\right)
(\rho_{m}+ \rho_\Lambda + p_{\Lambda}) = 0.
\end{equation}


\section{Solutions of the field equations}

Now, we solve the field equations Eqs. (\ref{7})- (\ref{10}), which are a system of four independent field equations in six unknowns 
parameters $a_{1}, a_{2}, \rho_{\Lambda}, \rho_{m}, p_{\Lambda} $ and $\phi$. Therefore, to obtain explicit solutions of the system, 
 For solving the above highly non-linear differential equations we have required the following physically significant conditions, which 
are relating to these parameters.\\

We assume that the expansion scaler ($\theta$) is proportional to shear scalar ($\sigma$). The condition leads to 
$\frac{\dot{a_{1}}}{a_{1}}=m\frac{\dot{a_{2}}}{a_{2}}$, which yields \cite{ref54}
\begin{equation}
\label{12} a_{1} = a_{2}^{m} ~.
\end{equation}
The motivation behind assuming this condition is given \cite{ref55}.\\

 The energy density of GGPDE $(\rho_{\Lambda})$ defined by \cite{ref42}\\
\begin{equation}
\label{13} 
\rho_{\Lambda} = {M^2_{pl}}(\alpha_1 H +\alpha_2 H^2)^{\kappa},
\end{equation}
where $H$ is the Hubble parameter, ${M^2_{pl}}$ = $\frac{1}{8\pi G}$ is the reduced Plank mass,  $\kappa$ is PDE parameter, and 
$\alpha_{1}$, $\alpha_{2}$ are two arbitrary dimensionless parameters. \\

Motivated by the recent observational data, we consider the deceleration parameter $q$ as a linear function of Hubble parameter 
$H$ as \cite{ref56}   
\begin{equation}
\label{14} 
q = -\frac{a \ddot{a}}{\dot{a}^{2}} ~ = ~ \alpha + \beta H ~.
\end{equation}
Here $\alpha$ and $\beta$ are arbitrary constants. 

%
%
%
%

Solving Eq. (\ref{14}), $\alpha = -1$, the various authors \cite{ref56,ref57,ref58,ref59} obtained the solution as:
\begin{equation}
\label{15} a = \exp{\left(\frac{1}{\beta}\sqrt{2\beta t + k}\right)},
\end{equation}
where $k$ is an integrating constant. Recently this Eq. (\ref{15}) is used by many authors \cite{ref56,ref57,ref58,ref59}. \\

We have found the value of deceleration parameter from Eq. (\ref{19}) 
\begin{equation}
\label{16}
q=-1+\frac{\beta}{\sqrt{2\beta t +k}}.
\end{equation}
As it has been observed that the expansion rate of the universe is time dependent instead of being constant. So we will use $\alpha = -1$, 
for which we find the time dependent value of DP and not $\alpha \neq -1$ as it results into constant DP $q=-1$. \\

The average scale factor for the metric (\ref{6}) is define as $a = (a_1 {a_2}^{2})^{\frac{1}{3}} $.\\

Using the above relation and Eqs. (\ref{12}) and (\ref{15}), we get 
\begin{equation}
\label{17}
a_{1}=\exp{\left[\frac{3m}{(m+2)}\frac{1}{\beta}\sqrt{2\beta t +k}\right]}.
\end{equation}
Using above in Eq. (\ref{12}), we get 
\begin{equation}
\label{18}
a_{2}= \exp{\left[{\frac{3}{(m+2)}\frac{1}{\beta }\sqrt{2\beta t +k}}\right]}.
\end{equation}

Therefore, the geometry of the metric $(\ref{6})$ comes out to be 
\begin{equation}
\label{19}
ds^{2}=dt^{2}-\left(\exp{\left[\frac{6m}{(m+2)}\frac{1}{\beta}\sqrt{2\beta t +k}\right]}\right){dx^{2}}- 
\left( \exp{\left[{\frac{6}{(m+2)}\frac{1}{\beta }\sqrt{2\beta t +k}}\right]}\right) (dy^{2}+dz^{2}).
\end{equation}

Now from equations (\ref{10}), (\ref{17}) and (\ref{18}) the scalar field is obtained as:
\begin{equation}
\label{20}
\phi^{\frac{(n+2)}{2}}= \frac{(n+2)}{2}\left[\exp{\left( \frac{-3}{\beta}\sqrt{2\beta t +k}\right)} 
\left(\frac{-c_1 }{3} \sqrt{2\beta t +k} - \frac{\beta}{9}\right) +c_{2} \right],
\end{equation}

where $c_1$ and $c_2$ are constants of integration. \\

Now we find the energy density of GGPDE from (\ref{13}) as
 \begin{equation}
 \label{21}
 \rho_{\Lambda} = \left[\frac{1}{ (\sqrt{2\beta t +k})}
 \left(\alpha_{1} + \alpha_{2}\frac{1}{ \sqrt{2\beta t +k} } \right) \right]^{\kappa}.
 \end{equation}
 
 From Eqs. (\ref{9}), (\ref{17}), (\ref{18}), (\ref{20}) and (\ref{21}) we find the energy density of the matter as 
\begin{equation}
\label{22}
\rho_{m} = \frac{9(2m+1)}{(m+2)^2 ({2\beta t +k})} + 
\frac{\omega c_{1}^2}{2}\exp{\left[\frac{-6}{\beta}\sqrt{2\beta t +k}\right]} - \left[\frac{1}{ (\sqrt{2\beta t +k})}
\left(\alpha_{1} + \alpha_{2}\frac{1}{ \sqrt{2\beta t +k} } \right) \right]^{\kappa}.
\end{equation}
The expression for pressure of holographic dark energy is obtained as:
\begin{equation}
\label{23}
p_{\Lambda} = {\left(\frac{-3\beta(m+1)}{(m+2) ({2\beta t +k})^\frac{3}{2}} +\frac{9(m^2 +m+1)}{(m+2)^2 (2\beta t +k)} - 
\frac{\omega c_{1}^2}{2}\exp{\left[\frac{-6}{\beta}\sqrt{2\beta t +k}\right]}\right)}.
\end{equation}

From Eqs. (\ref{23}) and (\ref{21}), we obtain equation of state parameter of dark energy $\omega_\Lambda$
\begin{equation}
\label{24}
\omega_{\Lambda} = \frac{{\left(\frac{-3\beta(m+1)}{(m+2) ({2\beta t +k})^\frac{3}{2}} +\frac{9(m^2 +m+1)}{(m+2)^2 (2\beta t +k)} - 
\frac{\omega c_{1}^2}{2}\exp{\left[\frac{-6}{\beta}\sqrt{2\beta t +k}\right]}\right)}}{\left[\frac{1}{ (\sqrt{2\beta t +k})} 
\left(\alpha_{1} + \alpha_{2}\frac{1}{ \sqrt{2\beta t +k} } \right) \right]^{\kappa}}.
\end{equation}

The density parameter $\Omega$ is given by 
\begin{equation}
\label{25}
\Omega = \frac{\rho_{\Lambda} + \rho_{m}}{3 H^{2}},
\end{equation}
which is equal to
\begin{equation}
\label{26}
\Omega = {\left(\frac{3(2m+1)}{(m+2)^2 } + \frac{\omega c_{1}^2 }{6} (2\beta t +k)\exp{\left[\frac{-6}{\beta}\sqrt{2\beta t +k}\right]}\right)}. 
\end{equation}


\section{Experimental data and calculations of parameters}

From Eq. (\ref{14}), we have $q_0 = -1 + \beta H_0$ , where $q_0$ and $H_0$ are the present values of DP and Hubble parameter respectively. \\

Also, we have $H=\frac{\dot{a}}{a}=\frac{1}{\sqrt{2\beta t +k} }$, so that $H_0 =\frac{1}{\sqrt{2\beta t_0 +k} }$. Using $H_0.t_0=1$, we 
can get $H_0=\frac{1}{\sqrt{2\frac{\beta}{H_0} +k} }$. Solving it for $k$, we have $k = \frac{1}{H_0}[\frac{1}{H_0}-2\beta]$.\\

Here three cases has been considered which is based on three different data:\\

\textbf{Case I: Based on supernovae type la union data} \\

Now, taking $q_0 = -0.73$ and $H_0 = 73.8$ \cite{ref60}, we obtain $\beta = 0.003658$. For these values $\beta$ and $H_0$, we get $k= 0.000084$.\\

\textbf{Case II: Based on SN la data in combination with BAO and CMB observations}\\

Taking $q_0 = -0.54$ and $H_0 = 73.8$ \cite{ref61} , we obtain $\beta = 0.0062$. For these values $\beta$ and $H_0$, we get $k= 0.00001$.\\

\textbf{Case III: Based on current data in combination with OHD and JLA observations}\\

Taking $q_0 = -0.52$ and $H_0 = 69.2$ \cite{ref62}, we obtain $\beta = 0.0069$. For these values $\beta$ and $H_0$, we get $k= 0.0000083$.\\

\begin{table}[H]
	\centering
	{\begin{tabular}{@{}ccccc@{}}
			\hline\hline 
			$q_0$ & $H_0$ & $\beta$ & $k$ & Reference \\
			\hline\hline
			$-0.73$ & $73.8$  & $0.003658$ & $0.000084$ & \cite{ref60}\\ 
			\hline
			$-0.54$ & $73.8$  & $0.0062$ & $0.00001$ & \cite{ref61}\\ 
			\hline
			$-0.52$ & $69.2$  & $0.0069$ & $0.0000083$ & \cite{ref62}\\ 
			\hline
	\end{tabular}}
	\caption{For different observed values of deceleration parameter $(q_0)$, Hubble parameter $(H_0)$, values of ($\beta$, $k$).}
\end{table}

In this paper we have drawn all the figures for these different values of ($\beta$, $k$) as (0.003658, 0.000084), (0.0062, 0.00001), 
and (0.0069, 0000083) respectively.
\\

The Hubble parameter $H$, scalar expansion $\theta$, shear scalar $\sigma$ and the average anisotropy parameter $A_m$ are calculated for 
our model which is defined as:
\begin{equation}
\label{27}H=\frac{\dot{a}}{a} ~ = ~ \frac{1}{\sqrt{2\beta t +k} }
\end{equation}
\begin{equation}
\label{28}\theta~=~3H=~\frac{3}{\sqrt{2\beta t +k} }~~
\end{equation}
\begin{equation}
\label{29}\sigma^{2}=\frac{1}{2}\left[(H^2_{1}+ 2H^2_{2})-\frac{\theta^2}{3}\right]~=~{\frac{3(m-1)^{2}}{(m+2)^{2}{(2\beta t +k)}}} ~ 
\end{equation}
\begin{equation}
\label{30}
A_{m}=\frac{1}{3}\sum_{i=1}^{3}\left(\frac{\triangle H_{i}}{H}\right)^2={\frac{2(m-1)^{2}}{(m+2)^{2}}}
\end{equation}

\section{Results and discussions}
Figure $1(a)$ depicts cosmological model with DP ($q$) versus time $t$. It can be observe that for  $k$ = $0.000084$ and 
$\beta$ = $0.0036$, our models are only in accelerating phase, but for $k$ = $0.000012$ and $\beta$ = $0.0064$, the model shows a 
transition phase from +ve to -ve deceleration parameter $q$ i. e. our model is transforming from $ q > 0$ (deceleration) to 
 $q<0$ (acceleration) phases. Similarly for the current data $k= 0.0000083$ and $\beta = 0.0069$ models is decelerating to the 
accelerating phase transforming. Since DP ($q$) value lies between the range
$-1 < q < 0$ and the expansion of the current universe in accelerating are exposed by SNe 1a data. So 
the value of the decelerating parameter is consistent with recent observations. \\ 

The value of scale factor ($a(t)$ ) in the form of $z$ is $a(t)=\frac{a_0}{1+z}$,  where $a_0$ is the current value of scale factor. 
From Eq. (\ref{19}), we find $ ln(a) = \frac{\sqrt{2\beta t + c_{1}}}{\beta}$. And we also obtain the value of 
$ln(a)$ in term of redshift $z$, i.e. $ln(a)= ln(a_0)-ln(1+z)$ from $a(t)=\frac{a_0}{1+z}$. Using this value 
in Eq. (\ref{20}), we obtain $q(z)=-1+\frac{1}{ln(a_0)-ln(1+z)}$ .\\

For our derived model variation of redshift $z$ versus $t$ (cosmic time) is demonstrate by figure 1(b). In this figure we have seen that 
for this value $a_{0}$ redshift $z$ is a monotonic decreasing with increase of $t$. So it can be say that for our model derived $z$ 
starts with a small positive value at $t$ equal to $0$ and $z$ tends to $-1$ as $t$ tends to $\infty$. \\

Figure 1(c) shows $q-z$ plane for generalized $q$-parametrization given by $q(z)=-1+\frac{1}{ln(\frac{(a_0)}{(1+z)})}$ for three cases 
of  $(\beta, k)$ i.e.( $\beta$ = $0.003658$, $k$ = $0.000084$, $\beta$ = $0.0062$, $k$ = $0.00001$ and  $\beta$ = $0.0069$, $k$ = $0.0000083$ ) 
and we get the value of $a_{0}$. For plot the graph we use these values.\\ 

In the figure we have discovered that $q(z)$ expansion undergoes a smooth transition from a decelerated stage to accelerated stage and 
$q \to -1$ as $z \to -1$. Currently, \cite{ref63,ref64} have discovered the change redshift from decelerating to accelerating in 
modified gravity cosmology. SNe type Ia dataset has given the progress from past deceleration to ongoing acceleration. 
More recently, in 2004 the HZSNS team have found $z_{t} = 0.46 \pm 0.13$ at $(1\;\sigma)$ c.l. \cite{ref4}. 
It is improved in 2007 to  $z_{t} = 0.43 \pm 0.07$ at $(1\;\sigma)$ c.l. \cite{ref5}. SNLS
\cite{ref65} compiled by \cite{ref66}, provide a progress redshift $z_{t} \sim 0.6 (1\; \sigma)$ in improved 
agreement with the flat $\Lambda$CDM model ($z_{t} = (2\Omega_{\Lambda}/\Omega_{m})^{\frac{1}{3}} - 1 \sim 0.66$).\\

Moreover the reconstruction of $q(z)$ is done by the joined $(SNIa + CC + H_{0})$, which have obtained the transition redshift 
$z_{t} = {0.69}^{+0.09}_{-0.06}, {0.65}^{+0.10}_{-0.07}$ and ${0.61}^{+0.12}_{-0.08}$ within  $(1 \sigma)$  \cite{ref67}. which are seen
as well consistent with past outcomes \cite{ref68,ref69,ref70,ref71,ref72} including the $\Lambda CDM$ expectation $z_{t} 
\approx 0.7$. Another limit of transition redshift is $0.6 \leq z_{t} \leq 1.18$ ($2\sigma$, joint examination ) \cite{ref73}.

\begin{figure}[H]
	(a)\includegraphics[width=9cm,height=6.5cm,angle=0]{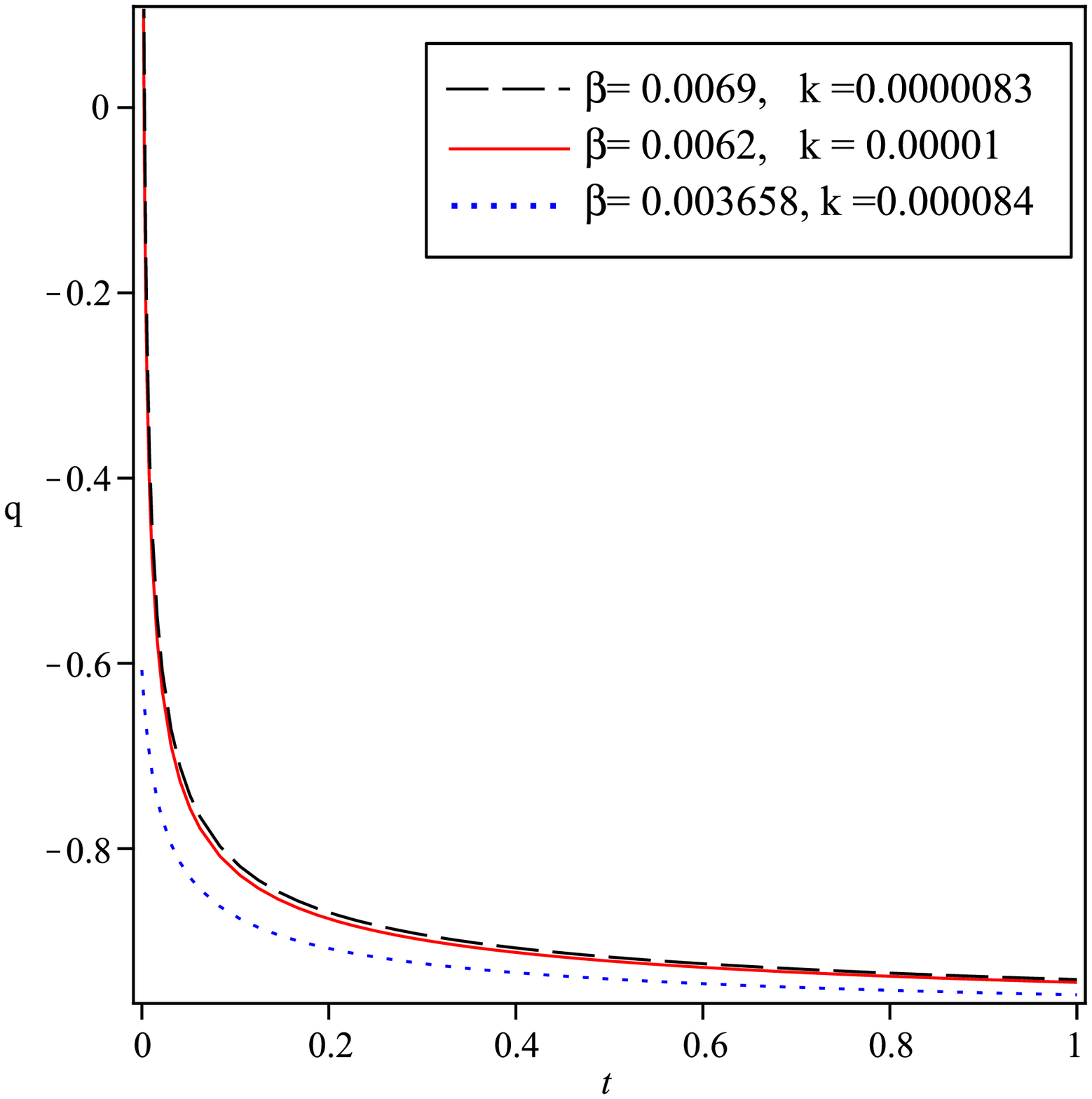}
	(b)\includegraphics[width=9cm,height=6.5cm,angle=0]{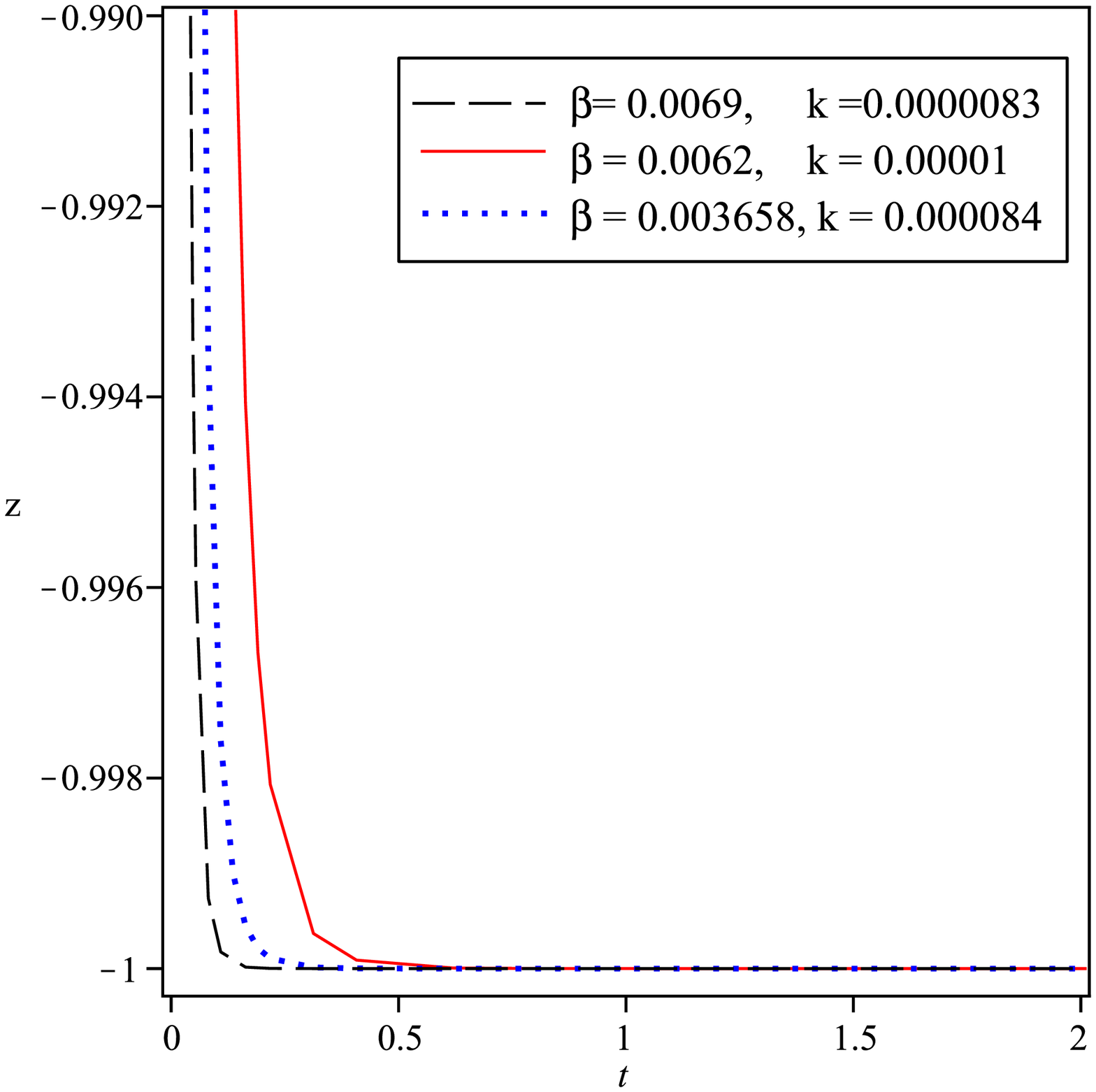}
	\\
	(c)\includegraphics[width=9cm,height=6.5cm,angle=0]{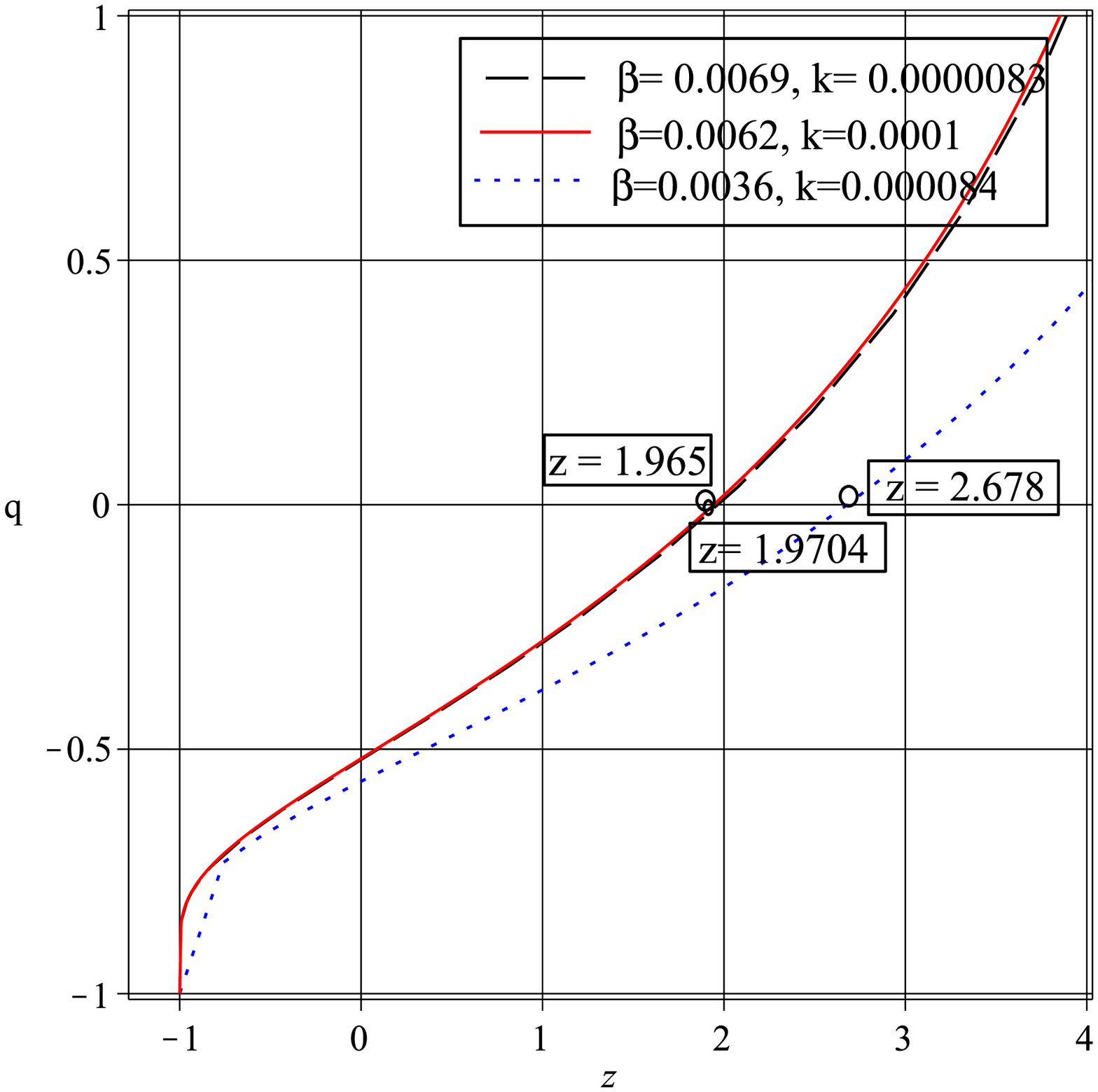}
	(d)\includegraphics[width=9cm,height=6.5cm,angle=0]{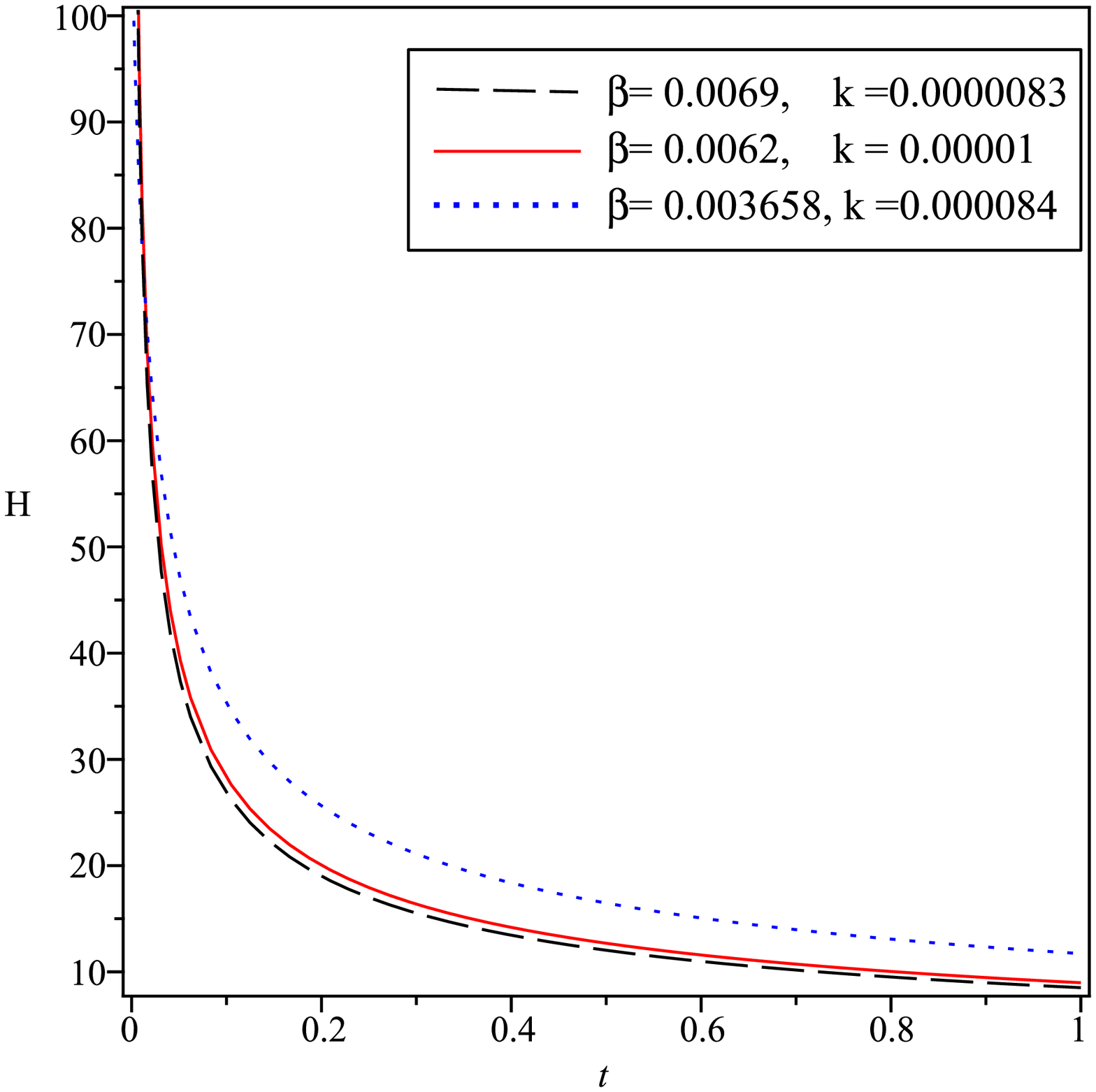}
	\caption{(a)The graph between $q$ versus  $t$, (b) The graph between $z$ versus  $t$,  
	(c)The graph between $q$ versus red shift $z$, (d) The graph of Hubble parameter $H$ versus cosmic time $t$.}
\end{figure}
 From the figure it is clear that the transition redshift from decelerated to accelerated expansion occurs at 
 $z_{t} \cong 1.965$,  $z_{t} \cong 1.9704$ and $z_{t} \cong 2.678$ (see Fig. $1c$) for above three cases which are found to be well consistent 
 with the Type Ia supernovae Hubble diagram, including the farthest known supernova SNI997ff at $z \approx 1.7$ \cite{ref74,ref75}). 
 \\
 
Figure $1(d)$ depicts between the Hubble parameter and time. We observed that
$H \to 0$ as $t \to \infty$ as per theoretical desire.

\begin{figure}[H]
	(a)\includegraphics[width=9cm,height=7cm,angle=0]{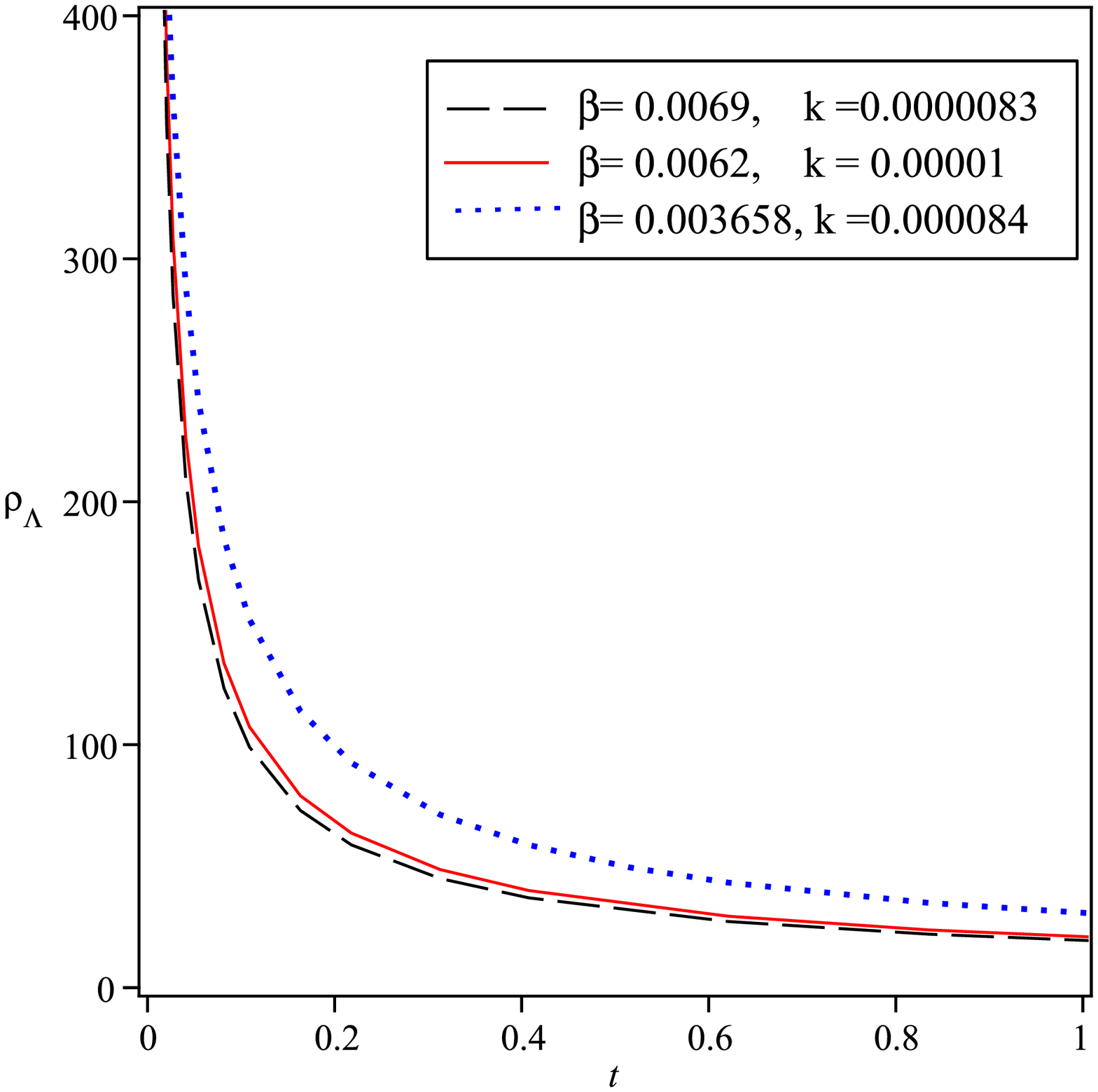}
	(b)\includegraphics[width=9cm,height=7cm,angle=0]{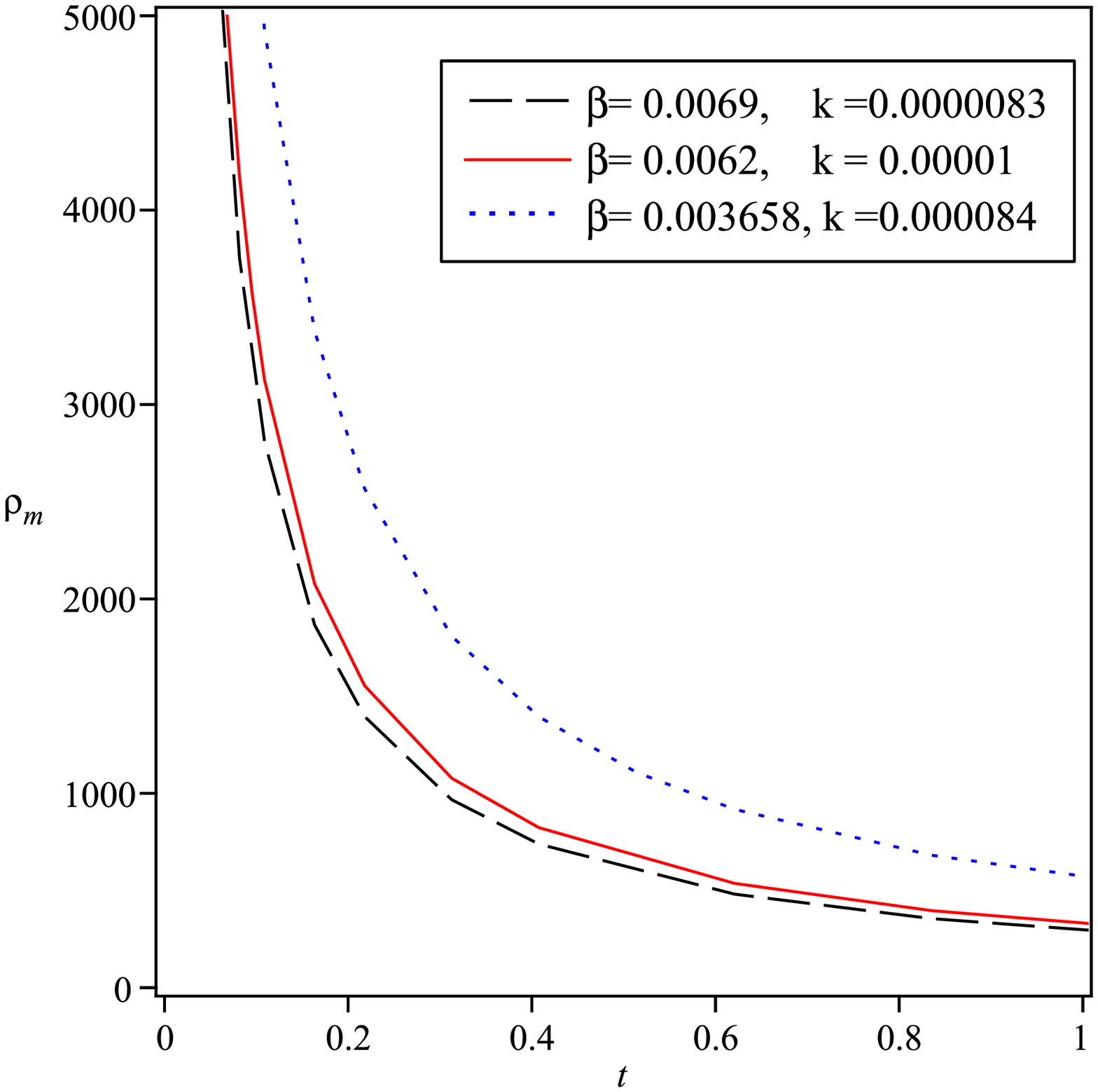}
	\caption{(a)The plot of  energy density of GGPDE versus time $t$, (b) The plot of energy density of 
	matter versus cosmic time $t$ .}
\end{figure}

We observe that from Eq. (\ref{25}) and it's corresponding figure $2(a)$, shows that for all three cases $\rho_{\Lambda}$ remains 
positive during the cosmic evolution. Similarly from Eq. (\ref{26}) and it's corresponding figure $2(b)$, it indicates that the energy 
density of matter $\rho_{m}$ decreases with increase of time and it also remains positive for all values of $k$ and $\beta$ during 
the cosmic evolution. We can see first it decreased sharply and then gradually and approach to a small 
positive values at the present epoch.  $\rho_{\Lambda}$ and $\rho_{m}$ tends to $0$ as $t \to \infty$. Since decrease in density 
implies the increase in volume i.e. our derived model represents the expansion 
of the universe.  


\begin{figure}[H]
	(a)\includegraphics[width=9cm,height=7cm,angle=0]{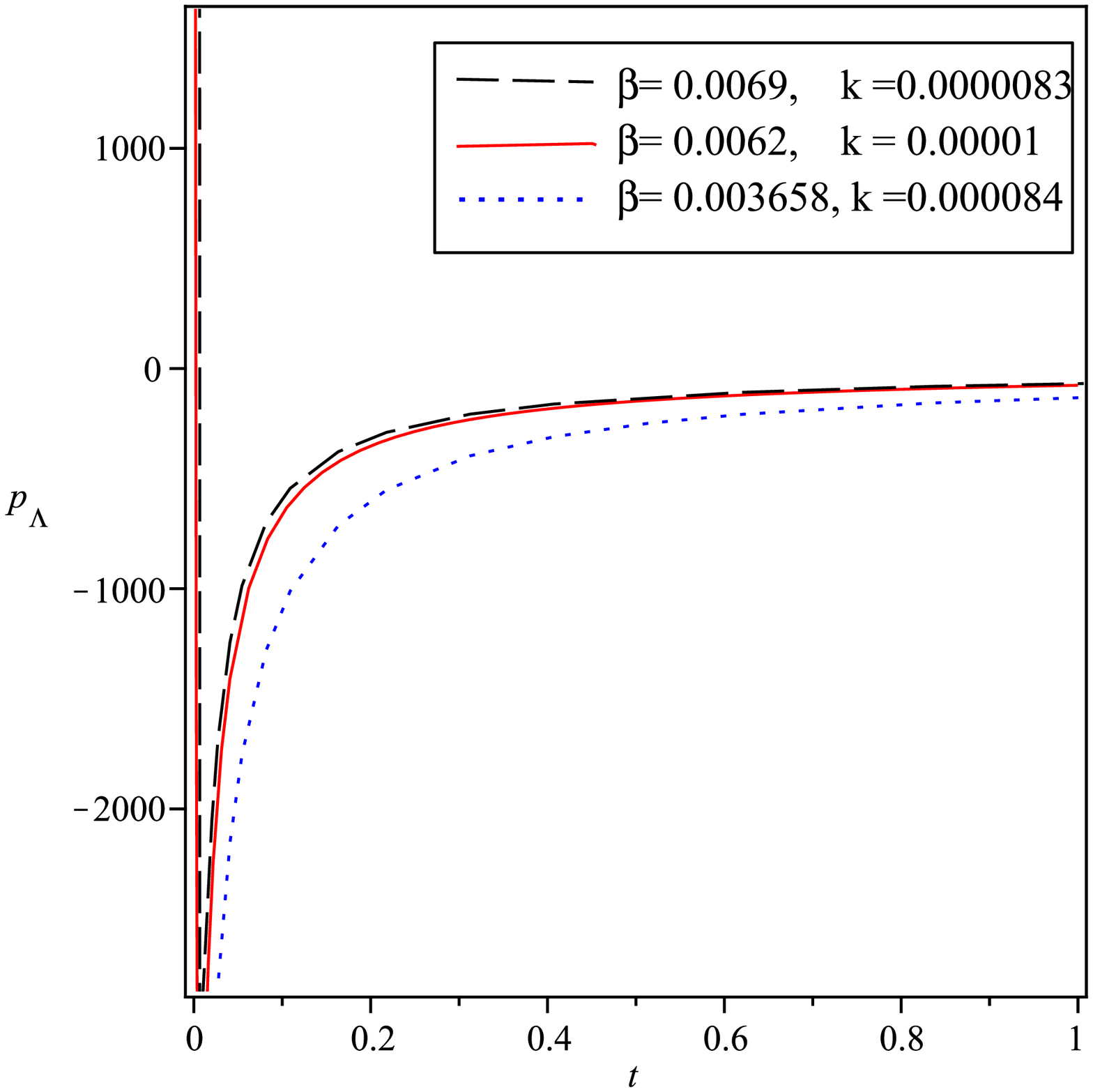}
	(b)\includegraphics[width=9cm,height=7cm,angle=0]{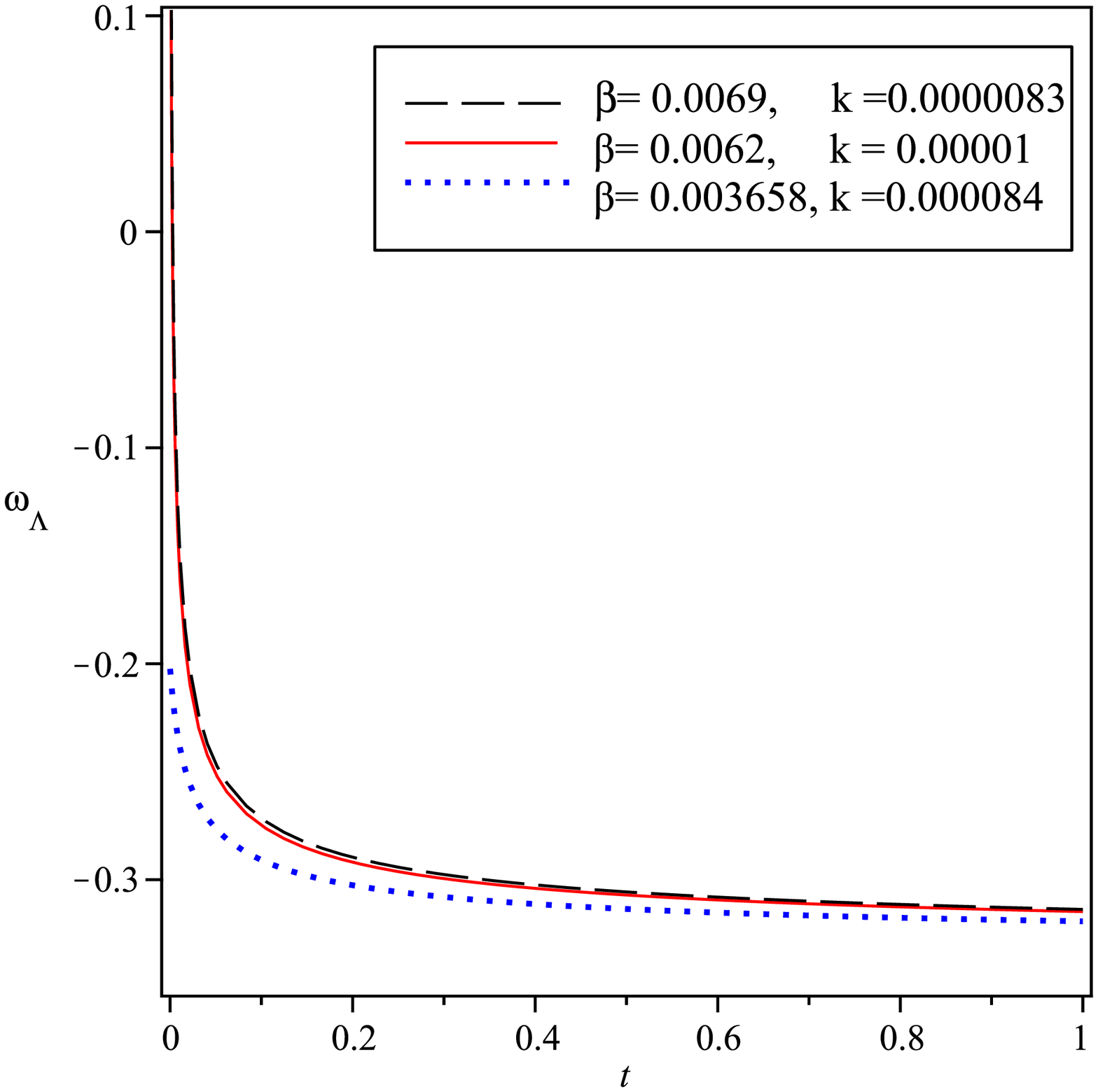}
	\caption{(a)The plot  of pressure $p_\Lambda$ of GGPDE versus time $t$, (b) The plot of equation 
	of state parameter $\omega_\Lambda$ versus cosmic time $t$ }
	\end{figure}

Figure $3(a)$ depicts pressure $p_{\Lambda}$ versus time $t$. It is observed that the value of isotropic pressure was positive in the 
beginning. Later it decreased rapidly and then it remained negative throughout evolution. This negative pressure put the equation of state 
$\omega=\frac{p_\Lambda}{\rho_\Lambda}<0$, as prescribed by recent observations and it may be a cause of the accelerated expansion of 
the universe. Figure $3(b)$ demonstrate the equation of state parameter $\omega$ for all three observational value. \\

From Eq. (\ref{28}) and its corresponding figure $3(b)$, it is clearly shown that for all three data firstly EoS
parameter $\omega_\Lambda$ was positive then decreased sharply and approach to a small negative value at the present epoch. The EoS 
parameter of DE may begin in phantom $\omega\le{-1} $ or quintessence $\omega\ge{-1}$ region and tends to $-1$ by 
exhibiting various patterns as $t$ increases. Here $\omega_{\Lambda}$ is less than $-1$ so there is a phantom region. i.e. 
$\omega_{\Lambda}$ evolves with negative and its range is in nice agreement with large scale structure data  \cite{ref76}.\\
\begin{figure}
	\centering
	\includegraphics[width=9cm, height=7cm]{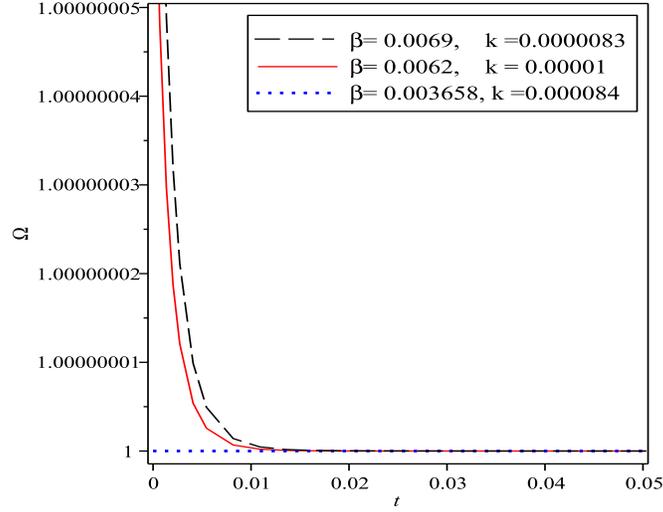}\\
	\caption{Total energy density $(\Omega)$ of GGPDE versus time $t$. }
\end{figure}

Figure $4$ shows the graph of $\Omega$ versus $t$ for all three values of ($\beta, k$). We observe that $(\Omega)$ approaches close to $1$ as 
$t \to \infty$ i.e. the total energy density of GGPDE models at present is $1$. \\

Jerk is the rate of change of acceleration; that is, the time derivative of acceleration. Many authors \cite{ref77, ref78} have indicated 
the significance of jerk as a tool to reconstruct the cosmological models. Recently,  \cite{ref79,ref80} have discussed the parametrization 
of time evolving jerk parameter models.\\

\begin{figure}[H]
\centering
\includegraphics[width=8cm, height=7cm]{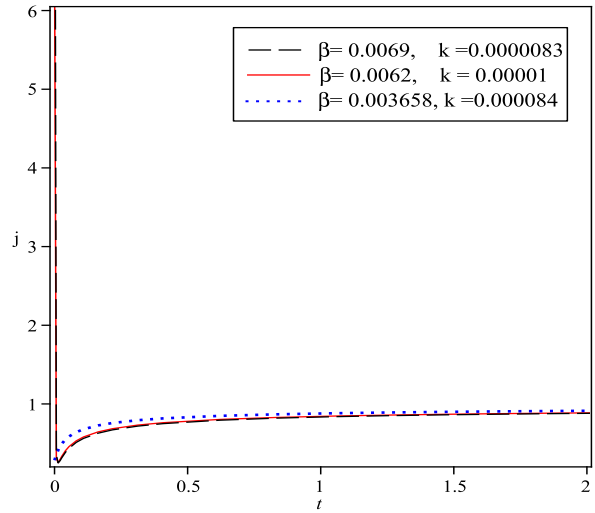}\\
\caption{Jerk parameter of GGPDE versus time $t$. }
\end{figure}

We hereby define the cosmic Jerk parameter such as:
	
\begin{equation}
	\label{31}
	j= -\frac{1}{aH^{3}}\left(\frac{d^{3}a}{dt^{3}}\right)= q +q^{2} - \frac{\dot{q}}{H}
	\end{equation} 
	\begin{equation}
	\label{32}
	j= \frac{3 \beta^{2}}{2\beta t +k} - \frac{3\beta}{\sqrt{2\beta t +k} } +1
	\end{equation}
	Figure $5$ demonstrate that firstly the graphs decrease rapidly for all three value $\beta = 0.0069$ \& $k = 0.0000083$; $\beta = 0.0062$ 
	\& $k = 0.00001$, and $\beta = 0.0069$ \& $k = 0.0000083$  then approach to $1$ through the evolution history i.e. the best suitable value 
	of $j$ is very close to $1$ is showed by this figure and it clearly indicates that the model is	extremely close to $\Lambda $CDM model. 
	A well known constrained value of jerk parameter is found by statistical analysis with different observational data sets like BAO (baryon 
	acoustic oscillation data), SNe (Ia supernova data) and OHD (observational Hubble parameter data). A well known constrained value of 
	jerk parameter is found by statistical analysis with different observational data sets like BAO (baryon 
	acoustic oscillation data), SNe (Ia supernova data) and OHD (observational Hubble parameter data)
	and the model remains at very close proximity of the $\Lambda$CDM. 

  \section{Squared sound speed}
 Many authors \cite{ref81,ref82} have examined the speed of sound constraints of dynamic DE models with EoS parameter varying in 
 time and concluded that the speed of the sound constraint of DE is very low. If the speed of sound lies between zero and 1, the system
 is stable.\\

 Now for any fluid, the adiabatic square sound speed is described as ${v_s}^2= dp/d \rho$. Besides the positivity of ${v_s}^2$, the 
 condition of causality must satisfied. Causality means the speed of sound should not surpass the speed of light. Since we use 
 relativistic units where $c$= $G$= $1$, the speed of sound should be within the range
  $0 \leq  dp/d\rho \leq 1 $.
   
  \[ {v_s}^2 = \frac{{\left(\frac{-9\beta^{2}}{(2m+1) ({2\beta t +k})^\frac{5}{2}} +\frac{18 \beta}{(2m+1)^2 (2\beta t +k)^\frac{3}{2}} - 
  \frac{\omega c_{1}^2 (1+\frac{2}{m})}{\sqrt{2\beta t +k}}\exp{\left[\frac{-2(m+2)}{m}\frac{1}{\beta}\sqrt{2\beta t +k}\right]}\right)}}
  { \bigg[ {\kappa} \left( \frac{1}{ (\sqrt{2\beta t +k})}
  	\left(\alpha_{1} + \alpha_{2}\frac{1}{ \sqrt{2\beta t +k} } \right) \right)^{\kappa} \left[ \frac{1}{ (\sqrt{2\beta t +k})}
  	\left(\alpha_{1} + \alpha_{2}\frac{1}{ \sqrt{2\beta t +k} } \right) \right] \times } 
  \]
  \begin{equation}
  \label{33}
  \left( \frac{- \beta}{ ({2\beta t +k})}
  \left( \alpha_{1} + \alpha_{2}\frac{1}{ \sqrt{2\beta t +k} } \right) 
  -\frac{\alpha_{2} \beta}{(2\beta t +k)} 
  \right) \bigg] .
  \end{equation}
  \begin{figure}[H]
  	\centering
  	(a)\includegraphics[width=7cm,height=5cm,angle=0]{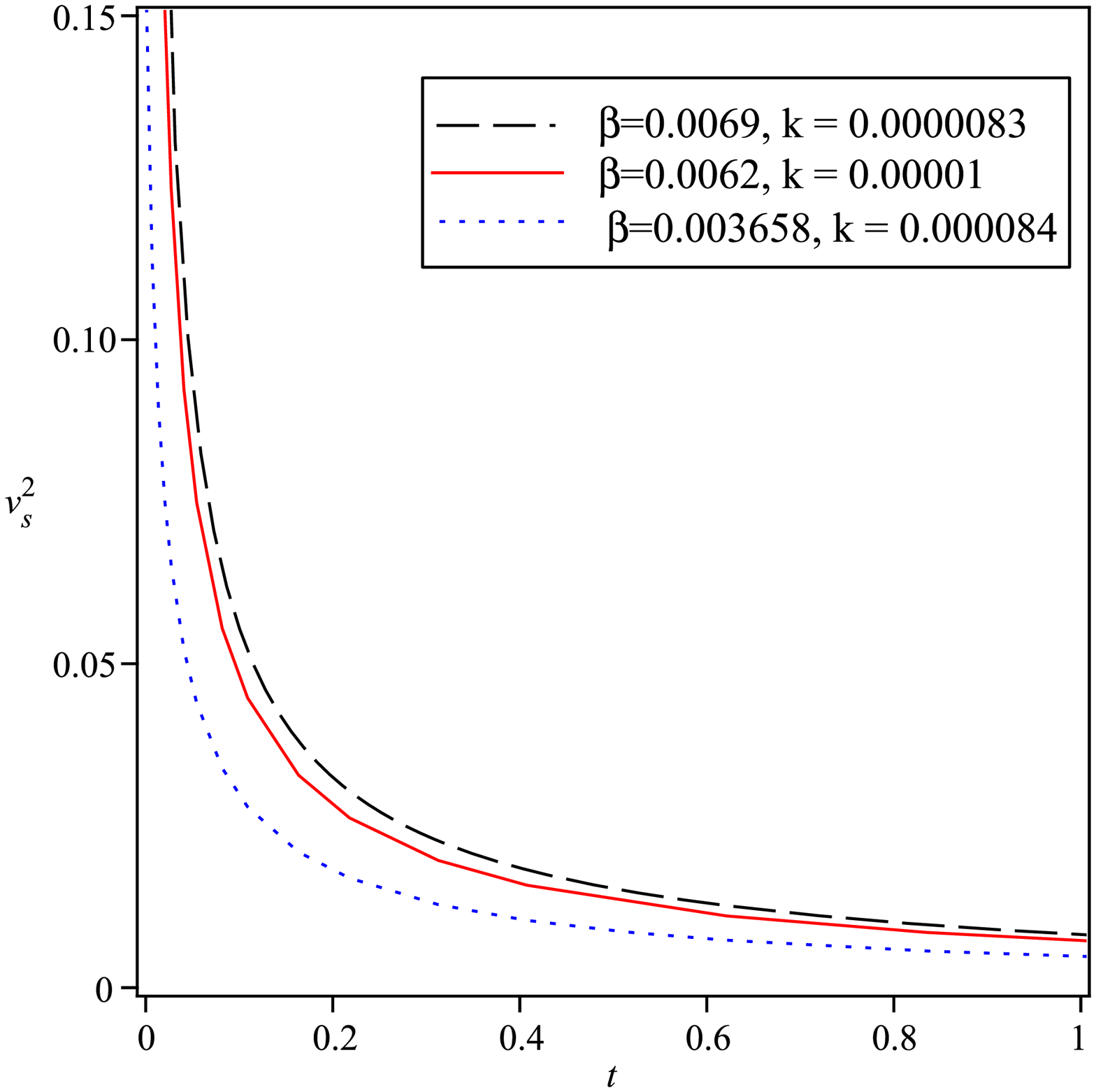}
  	\caption{The graph between $v^{2}_s$ versus time $t$ is shown}
  \end{figure}
  Figure $(6)$  depict the plots of speed of sound with time for all three 
  values of $\beta$ and $k$. We observe that $v_{s} < 1$ throughout the
   evolution of the universe. From the figure, it is sorted out that 
   throughout the universe evolution sound speed remains less than light velocity ($c=1$). 
   
  \section{The $\omega_D - \omega_D^{'}$ analysis}
   Caldwell and Linder \cite{ref83} implemented a plane analysis of  $\omega_D - \omega_D^{'} $. This analysis is a utile shaft 
   for separating vDE models by trajectories on their planes. The method is also applied to the quintessence model that leads to two forms 
 of its plane, i.e. the area belongs to the $(\omega_D <0, \omega _D^{'} > 0)$
  region refers to the thawing region while the area below the $(\omega_D < 0, \omega_D^{'} < 0)$ region indicates the freezing region. 
  Later this analysis was extended by many researchers to other dynamic DE models \cite{ref84,ref85,ref86,ref87}. Clearly, a fixed point 
  at$(-1, 0)$ is the $\Lambda$CDM model in the $\omega_D$ phase space. To discuss GGPDE model, here we use this analysis.\\
   On differentiating  Eq. (\ref {24}) w.r.t. $lna$ and find the value of 
 $\omega'_D$ as 

  \[ \omega'_D = \bigg[ \frac{{\left(\frac{-9\beta^{2}}{(2m+1) ({2\beta t +k})^\frac{5}{2}} +\frac{18 \beta}{(2m+1)^2 (2\beta t +k)^\frac{3}{2}} - 
  \frac{\omega c_{1}^2 (1+\frac{2}{m})}{\sqrt{2\beta t +k}}\exp{\left[\frac{-2(m+2)}{m}\frac{1}{\beta}\sqrt{2\beta t +k}\right]}\right)}}
  {\left[\frac{1}{ (\sqrt{2\beta t +k})}
  	\left(\alpha_{1} + \alpha_{2}\frac{1}{ \sqrt{2\beta t +k} } \right) \right]^{\kappa}} - 
  \]
  \[ {\kappa} \frac{{\left(\frac{3\beta^{2}}{(2m+1) ({2\beta t +k})^\frac{3}{2}} -\frac{9}{(2m+1)^2 (2\beta t +k)} + 
  \frac{\omega c_{1}^2 }{2}\exp{\left[\frac{-2(m+2)}{m}\frac{1}{\beta}\sqrt{2\beta t +k}\right]}\right)}}
  {\left[\frac{1}{ ({2\beta t +k})^\frac{\kappa}{2}}
  	\left(\alpha_{1} + \alpha_{2}\frac{1}{ \sqrt{2\beta t +k} } \right) \right]^{\kappa + 1}} \times
  \]
  \begin{equation}
  \label{34}
  \left( \frac{- \beta}{ \sqrt{2\beta t +k}}
  \left( \alpha_{1} + \alpha_{2}\frac{1}{ \sqrt{2\beta t +k} } \right) 
  -\frac{\alpha_{2} \beta}{\sqrt{2\beta t +k}} 
  \right) \bigg] .
  \end{equation}
  \begin{figure}[H]
  	\centering
  	\includegraphics[width=7cm,height=6cm,angle=0]{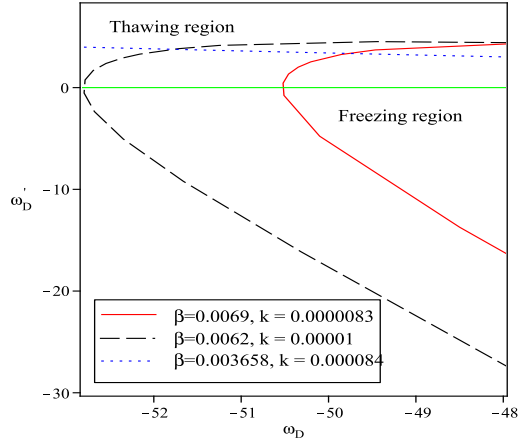}
  	~~\caption { The plot of evolutionary trajectories of the $\omega_D - \omega'_D$}
  \end{figure}
  We constructed the $\omega_D-\omega'_D$ plane by plotting $ \omega'_D$ versus $\omega_D$ as shown in Fig. $7$.
   It can be observed that the curves lead to both freezing and thawing regions for three values of parameter$\beta$ and $k$. Nevertheless, 
   the trajectories differ for the supernova Ia union data in the thawing region and for the BAO $\&$ CMB observations and OHD $\&$ JLA 
   observation curves correspond to both the freezing and thawing areas. Nevertheless, our model's trajectories differ mostly in the 
   freezing zone, as indicated by observational data (the expansion of the universe is comparatively more accelerated in the freezing area). 
 \section{Correspondence of quintessence field in GGPDE}  
 
 In the presence of non-relativistic matter represented by a barotropic perfect fluid, let us consider quintessence. The quintessence field 
 action is given in \cite{ref88}
 \begin{equation}
 \label{35}
 S = \int \sqrt{-g} d^4 x \left[ \frac{1}{2} M^2_{pl} R - \frac{1}{2} g^{ij} \phi_{;i} \phi_{;j} - V(\phi) \right]+ S_m ,
 \end{equation}
 
where $M_{pl}$ is the reduced Planck mass, $g^{ij}$ is the metric whereas $g$ is the metric determinant and $R$ is the scalar Ricci. In the 
above, $S_m$ is the action standard field, $V (\phi)$ is the potential for the quintessence field $\phi$ and $\phi{; i}$ is the covariant 
derivative of $\phi$. \\

By varying the action given in Eq. (\ref{35}) with respect to the metric and $\phi$ respectively we obtain, 
  \begin{equation}
  \label{36}
  3 M^2_{pl} H^{2} = \rho_m + \frac{\dot {\phi}^{2}}{2} +  V(\phi) , 
  \end{equation}
   \begin{equation}
   \label{37}
   M^2_{pl}( 2 \dot {H} + 3 H^{2}) = - \frac{\dot {\phi}^{2}}{2} +  V(\phi). 
   \end{equation}
   On solving Eqs. (\ref{36}) and (\ref{37}), we obtain potential versus cosmic time $t$. 
   
     \begin{equation}
     \label{38}
   V(\phi) = \frac{M^2_{pl}}{2} \left( \frac{-2 \beta}{(2 \beta t +k)^\frac{3}{2}} + \frac{6}{(2 \beta t +k)}\right) - \frac{1}{2} 
   {\left[\frac{\alpha_{1}}{ (\sqrt{2\beta t +k})} + \frac{\alpha_{2}}{ {2\beta t +k} }  \right]^{\kappa}}, 
     \end{equation}
     
     and scalar field  versus time $t$ is 
    \begin{equation}
    \label{39}
    \dot{\phi}^2 = \frac{6 M^2_{pl}}{(2 \beta t +k)} - \frac{3}{2} {\left[\frac{1}{ (\sqrt{2\beta t +k})}
    	\left(\alpha_{1} + \alpha_{2}\frac{1}{ \sqrt{2\beta t +k} } \right) \right]^{\kappa}} + \frac{M^2_{pl}}{2} 
    	\left( \frac{2 \beta}{(2 \beta t +k)^\frac{3}{2}} - \frac{6}{(2 \beta t +k)}\right), 
    \end{equation} 
  \begin{figure}[H]
  	(a)\includegraphics[width=9cm,height=7cm,angle=0]{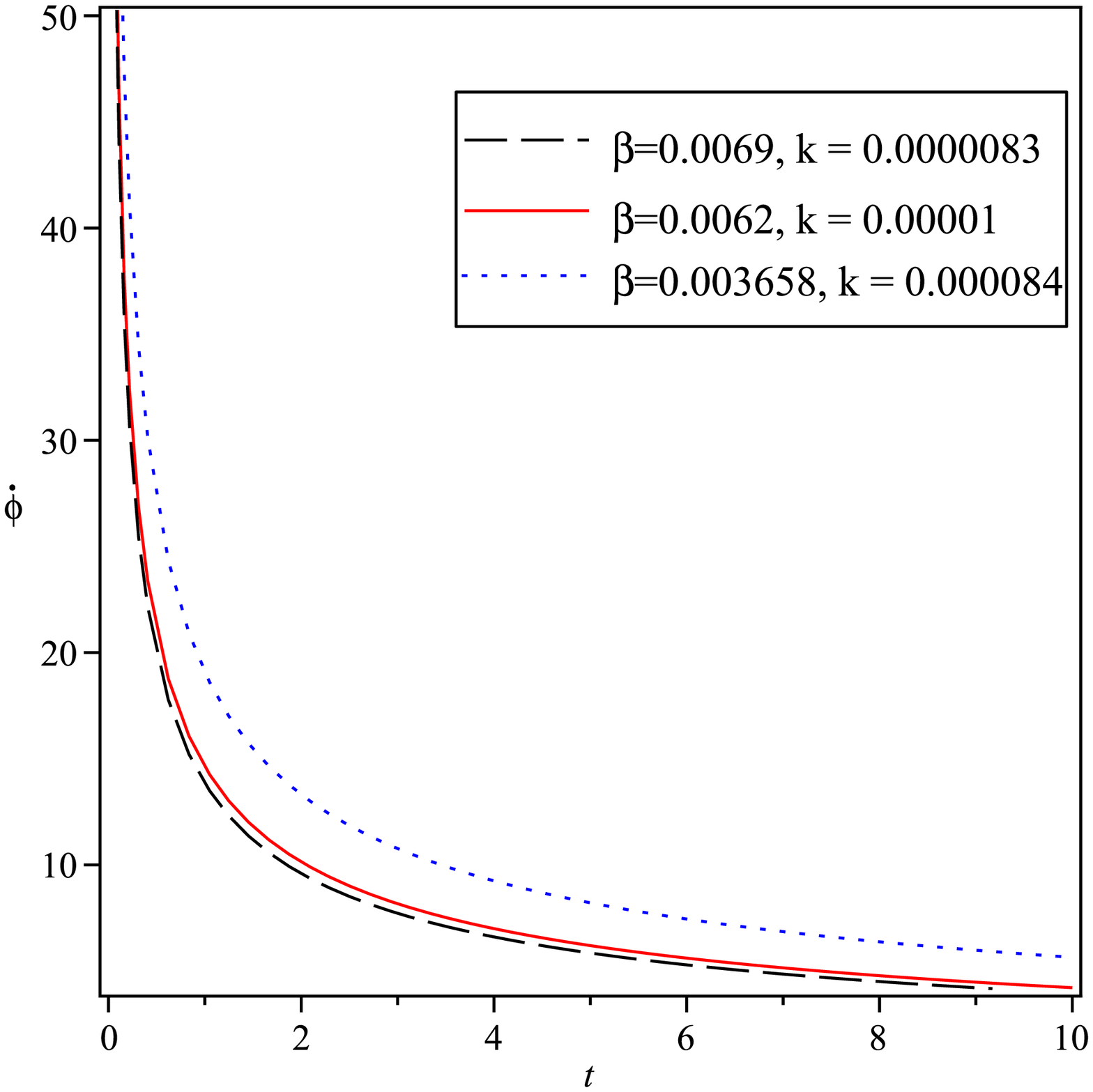}
  	(b)\includegraphics[width=9cm,height=7cm,angle=0]{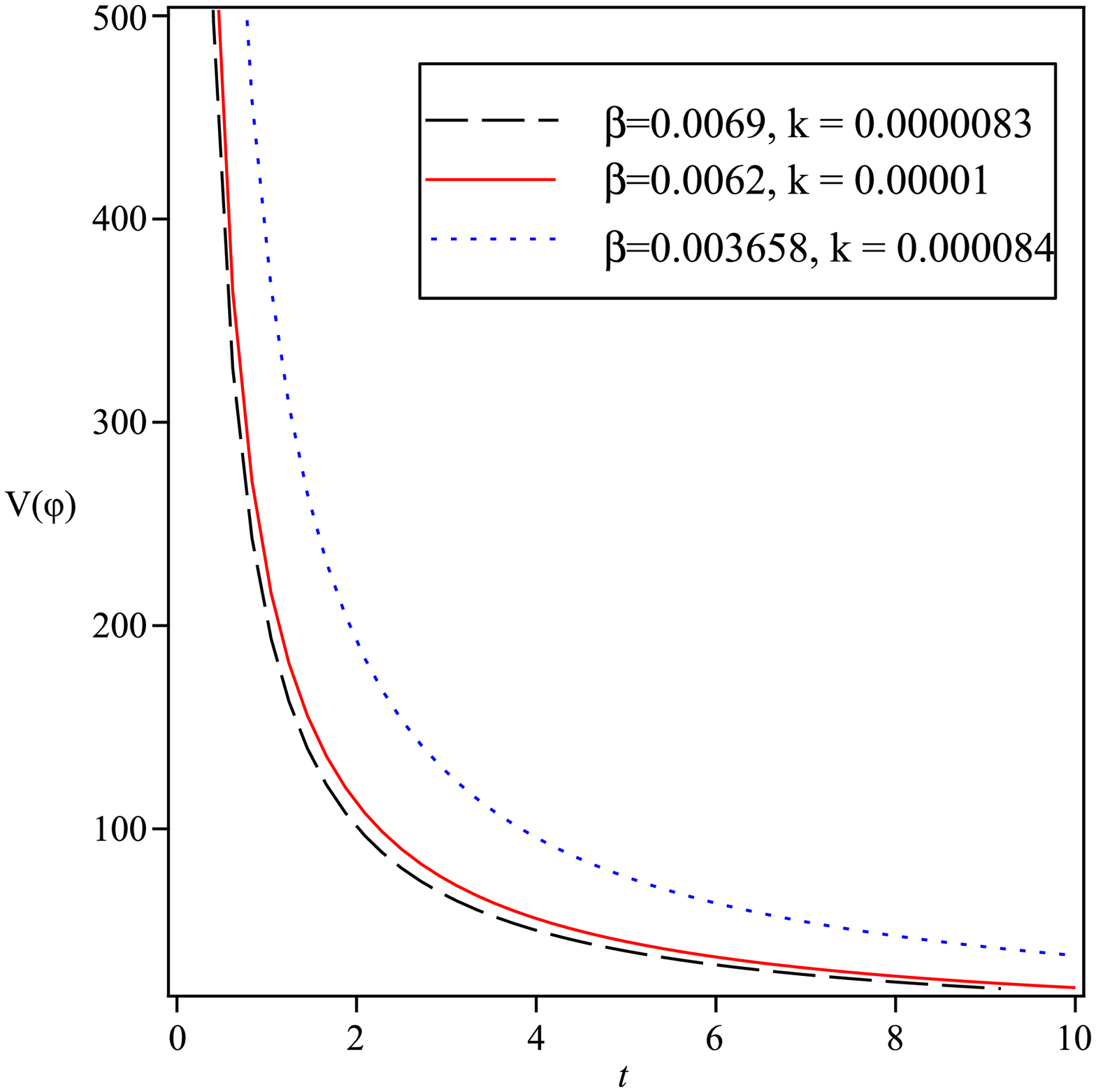}
  	\caption{(a)The plot of left hand sides of scalar field  versus time $t$, (b) The plot of potential versus cosmic time $t$.}
  \end{figure}
A canonical scalar field $\phi$ describe the quintessence model.
 A slowly varying potential of scaler field $V(\phi)$ can lead to the accelerating Universe.\\

 This method is close to slow inflation in the early Universe. But the difference is that it is not possible to ignore non-relativistic matter
 (dark matter and baryons) to accurately explain the nature of DE.\\

 Figure $(8a)$ shows the graph of scalar field $\phi$ with time $t$ for all three values of $\beta$ and $k$. Scalar field $\phi$ decreases 
 through the evolution of time and tends to small positive value at the present epoch. Figure $(8b)$ depicts potential of scalar field that 
 correspond to LRS Bianchi type-I GGPDE model, which is also decreasing function of time

\section{Concluding remarks}

A new LRS Bianchi type-I generalized ghost pilgrim DE anisotropic universe with time dependent DP $q$ and time dependent EoS 
parameter $\omega$ has been explored by new idea.

Here we have considered two assumptions: (i) $\sigma^{2}$ (shear scalar) $\propto$ $\theta$ (scalar expansion) and (ii) the DP $q$ as 
a linear function of $H$ ( Hubble parameter).\\

There are following main features of the model: \\
\begin{itemize}

 \item For different purposes, like in analysis of jerk parameter, correspondence with various scalar field models, thermodynamics laws, the 
evolution of our universe by extracting different cosmological parameters, the generalized ghost dark energy model play a very important 
role. In different modified theories, various cosmological aspects of this model have been also addressed. It should be mentioned here 
that $\rho_m$ (energy density of matter) and $\rho_\Lambda$ (the proper energy density), are positive in the deduced model. 

\item The anisotropic parameter $A_{m}$ is a constant and is non-zero for $m \neq 1$. ie. our model is anisotropic and may attain 
isotropy if $m=1$.

\item The interacting scenario of generalized ghost dark energy with CDM has been studied and cosmological parameters like jerk parameter, 
total energy density and equation of state parameter (EoS) have been evaluated by us. For three different 
well-known values of $k$ and $\beta$, equation of state parameter (EoS) of PDE parameter has been explored. 
$\omega_\Lambda$ is in the phantom region of our universe for all three observational values of $k$ and $\beta$.

\item The GGPDE gives the dynamics of  $\Omega$ given by Eq. (\ref {30}). In this derived model, $\Omega$ is 
obtained as time dependent which is a  positive decreasing with time and approach to $1$ at present epoch (Figure $4$).

\item It is observed that the best suitable value of jerk parameter is very close to $1$ at present time (Figure $5$). This shows that 
our GGPD model is very close to the $\Lambda$CDM model. 
 
 \item The model represents that $H$, $\sigma^{2}$ and $\theta$ diverge at $t = -\frac{k}{2\beta}$), but the volume becomes unity at 
 $t = -\frac{k}{2\beta}$). Hence, the GGPD model starts expanding with a big bang singularity at $t = -\frac{k}{2\beta}$). It may be observed 
 that when $t$ tends to  $\infty$, the volume becomes $\infty$. 
 
 \item The GGPD model represents an expanding, shearing, non-rotating and transit (from decelerating to accelerating) Universe.

\item The squared sound speed trajectories demonstrate positive behavior throughout the evolution and model 
stability are also discussed in (Fig.$6$). 

\item For our GGPD model, we also developed the $\omega_D-\omega_D^{'}$ plane in (Fig.$7$) correspond to both  thawing 
regions and freezing regions.\\. 

\item Nevertheless, our model's trajectories differ mostly in the freezing region, as indicated by observational data (the expansion of the 
universe is comparatively more accelerated in the freezing region). 

\item We also observed the quintessence correspondence with a scalar field $\phi$ and potential $V(\phi)$ in (Fig.$8a, 8b$).
For many different potentialities, the dynamics of quintessence in the presence of non-relativistic matter have been studied in 
detail \cite{ref83,ref88,ref89,ref90,ref91}. This result is 
a good agreement with the current scenario of the universe.

\end{itemize}

Hence, our constructed model and their solutions are physically acceptable. Therefore, for the better understanding of the 
characteristics within the framework of consistency GGPD models, the solution demonstrated in this paper may be helpful.


\end{document}